\documentstyle[manuscript,aps]{revtex}
\begin{document}
\draft
\title{Remarks on the Bergmann--Thomson Expression on Angular Momentum in
General Relativity}  
\author{Janusz Garecki}
\address{Institute of Physics, University of Szczecin, Wielkopolska 15, 70--451
Szczecin, POLAND} 
\date{\today}
\maketitle
\begin{abstract}
Often it is asserted that only by using of the symmetric Landau--Lifschitz
energy--momentum complex one is able to formulate a conserved angular momentum
complex in General Relativity ({\bf GR}). Obviously, it is an uncorrect
statement. For example, years ago, Bergmann and Thomson have given other, very
useful expression on angular momentum. This expression is closely joined to the
non--symmetric, Einstein canonical energy--momentum complex. In the paper we
review the Bergmann--Thomson angular momentum complex and compare it with that
of given by Landau and Lifschitz.  
\end{abstract}
\newpage
\section{Bergmann--Thomson expression on angular momentum in GR}
One can most easily obtain the canonical Bergmann--Thomson expression on total
angular momentum density, matter and gravitation, in the following way. 
At first, let us transform the Einstein equations written in mixed form and
multiplied by $\sqrt{\vert g\vert}$ 
\begin{equation}
\sqrt{\vert g\vert}G_i^k =
\beta\sqrt{\vert g\vert}T_i^k 
\end{equation}
to the so--called {\it superpotential form}
\begin{equation}
_E {K_i^{~k}} = {_F {U_i^{~[kl]}}}_{,l}, 
\end{equation}
where 
\begin{equation}
_E K_i^{~k} := \sqrt{\vert g\vert}\bigl(T_i^{~k} + _E t_i^{~k}\bigr)
\end{equation}
is the {\it canonical, Einstein's energy--momentum complex}, matter and
gravitation, and $_F {U_i ^{~[kl]}}$ mean {\it von Freud superpotentials}. 

From (2) we get, after series operations, the {\it Bergmann--Thomson
energy--momentum complex} $_{BT} K^{jk}$,  matter and gravitation. At first, we
form  
\begin{equation}
g^{ij}{} {_E K_i^{~k}} = g^{ij}{} {_F {U_i^{~[kl]}}}_{,l} = 
\bigl(g^{ij}{}{_F U_i^{~[kl]}}\bigr)_{,l} - {_F U_i^{~[kl]}}{}g^{ij}_{~~,l}
\end{equation}
or 
\begin{equation}
g^{ij}{}{_E K_i^{~k}} + {_F U_i^{~[kl]}}{}g^{ij}_{~~,l} = {_F U^{j[kl]}}_{,l}.
\end{equation}
Then, we write (5) in the form 
\begin{equation}
_{BT} K^{jk} = {_F U^{j[kl]}}_{,l},
\end{equation}
where 
\begin{equation}
_{BT} K^{jk} := _E K^{jk} + {_F U_i^{~[kl]}}{}g^{ij}_{~~,l} =: \sqrt{\vert
g\vert}\bigl(T^{jk} + _{BT} t^{jk}\bigr)
\end{equation}
is the {\it Bergmann--Thomson energy--momentum complex} [1,2,3,4] of  matter
and gravitation which satisfies local conservation laws 
\begin{equation}
{_{BT} K^{jk}}_{,k} = 0.
\end{equation}
Here $_{BT} t^{jk} \not= _{BT} t^{kj}$ mean the components of the so--called
{\it Bergmann--Thomson energy--momentum pseudotensor} of the gravitational
field [1,2,3,4]. Finally, from (6) we get 
\begin{eqnarray}
x^i {_{BT} K^{jk}} - x^j{_{BT} K^{ik}} & = & x^i {_F U^{j[kl]}}_{,l} - x^j {_F
U^{i[kl]}}_{,l} \nonumber \\
& = & \bigl(x^i{_F U^{j[kl]}} - x^j {_F U^{j[kl]}}\bigr)_{,l} - _F U^{j[ki]} +
_F U^{i[kj]},
\end{eqnarray}
 or 
\begin{equation}
x^i {_{BT} K^{jk}} - x^j {_{BT} K^{ik}} + _{BT} S^{ijk} =
M^{[ij][kl]}_{~~~~~~~,l},
\end{equation}
where 
\begin{eqnarray}
_{BT} S^{ijk} & := & _F U^{i[jk]} - _F U^{j[ik]} =  {\alpha\over\sqrt{\vert
g\vert}}\biggl[(-g)\bigl(g^{kj} g^{il} - g^{ki} g^{jl}\bigr)\biggr]_{,l}
\nonumber \\
& =: & {\alpha\over\sqrt{\vert g\vert}} g^{[ij][kl]}_{~~~~~~~,l}
\end{eqnarray}
and
\begin{equation}
M^{[ij][kl]} := x^i{_F U^{j[kl]}} - x^j {_F U^{i[kl]}}.
\end{equation}
The expression 
\begin{equation}
x^i {_{BT} K^{jk}} - x^j {_{BT} K^{ik}} + _{BT} S^{ijk} =:
_{BT} M^{ijk}
\end{equation}
is the {\it Bergmann--Thomson angular momentum complex} matter and gravitation
and the quantities $M^{[ij][kl]}$ are {\it superpotentials} [1].

The complex $_{BT} M^{ijk}$ satisfies local conservation laws 
\begin{equation}
_{BT} M^{ijk}_{~~~,k} = 0.
\end{equation}
One can interpret physically the angular momentum complex (13) as a sum of the
{\it orbital part}  
\begin{equation}
O^{ijk} := x^i {_{BT} K^{jk}} - x^j {_{BT} K^{ik}} = \sqrt{\vert
g\vert}\bigl(x^i T^{jk} - x^j T^{ik}\bigr) + \sqrt{\vert g\vert}\bigl(x^i
{_{BT} t^{jk}} - x^j {_{BT} T^{ik}}\bigr)
\end{equation}
of the angular momentum density of matter and gravitation (The matter part
includes also spin density [1]) and a {\it spinorial part} 
\begin{equation}
_{BT} S^{ijk} = _F U^{i[jk]} - _F U^{j[ik]} =
{\alpha\over\sqrt{\vert g\vert}} g^{[ij][kl]}_{~~~~~~~,l}
\end{equation}
of the gravitational angular momentum density.

We have from (6) 
\begin{equation}
2_{BT} K^{[ij]} = \sqrt{\vert g\vert}\bigl(_{BT} t^{ij} -
_{BT} t^{ji}\bigr) = {_{BT} S^{ijk}}_{,k} = \biggl({\alpha\over\sqrt{\vert
g\vert}}\biggr)_{,k} g^{[ij][kl]}_{~~~~~~~,l},
\end{equation}
because the dynamical energy--momentum tensor of matter $T^{ik}$ is symmetric:
$T^{ik} = T^{ki}$. 

The equality (17) justifies the above proposal of the physical interpretation
of the pseudotensor $_{BT} S^{ijk} = (-) _{BT} S^{jik}$ as a
quantity describing {\it canonical spin density}  of the gravitational field
\footnote{In Special Relativity the antisymmetric part of an
energy--momentum tensor is proportional to the ordinary divergence of a
quantity which describes canonical spin densities [5,6]. Here we follow this
line.}. 

It is very interesting that formal application of the
special-relativistic symmetrization procedure given by Belinfante (see,
e.g., [5,6]) to the complex $_{BT} K^{ij}$ with $S^{ijk} = _{BT}
S^{ijk}$ leads us immediately to the new, symmetric complex $_{BT}
K^{Nij} = \sqrt{\vert g\vert}\bigl(T^{ij} - {c^4\over 8\pi G}
G^{ij}\bigr)$ which trivially vanishes, i.e., it leads us to the Lorentz
and Levi-Civita solution of the energy-momentum problem for
gravitational field:  $_g T^{ik} = (-){c^4\over 8\pi G}G^{ik}$.

From (10) we get the following expression on the components $M^{ik} = (-)
M^{ki}$ of the global angular momentum of matter and gravitation for an 
isolated system endowed with an asymtotically Lorentzian
coordinates \footnote{The Bergmann--Thomson expression (10), likely as the
canonical Einstein energy--momentum complex, can be reasonably use only in the
case of an isolated system endowed with an asymptotically flat coordinates
(see, e.g., [2,3]).} 
\begin{equation}
_{BT} M^{ik} = {1\over c} \oint\limits_{over~ sphere~ having~ R\longrightarrow
\infty} \bigl(x^i{_F U^{j[0\alpha]}} - x^j {_F
U^{i[0\alpha]}}\bigr)n_{\alpha}r^2 d\Omega.
\end{equation}
$r^2 = x^2 + y^2 + z^2$;   $n_{\alpha}$ are the components of the unit
(exterior) normal to the sphere and $d\Omega = \sin\theta d\theta d\varphi$.

For the Schwarzschild spacetime equipped with an asymptotically Lorentzian
coordinates $(x^0 = ct,~x^1 = x,~x^2 = y,~x^3 = z)$ we get from (18) the
expected result 
\begin{equation}
_{BT} M^{ik} = 0.
\end{equation}
On the other hand, in the case of the stationary and axially symmetric Kerr's
spacetime (see, e.g., [7,8,9]) we obtain, in an asymptotically Lorentzian
coordinates also, that the only one component 
\begin{equation}
_{BT} M^{12} = (-) _{BT} M^{21}
= mac - {1\over 3}mac = {2\over 3}mac 
\end{equation}
of the $_{BT} M^{ik}$ is different from zero.

One should interpret the result (20) as giving the {\it global angular
momentum}, spinorial and orbital of matter and gravitation, for Kerr spacetime.
\section{Landau--Lifschitz expression on angular momentum density in
General Relativity and its comparison with Bergmann--Thomson expression}
One can easily  obtain the Landau--Lifschitz expression on angular momentum
density for  matter and gravitation in the following way [9]. At first, one
should transform the Einstein equations multiplied by  $(- g)$ 
\begin{equation}
(- g)G^{ik} = \beta (-g)T^{ik}
\end{equation}
to the {\it superpotential form} 
\begin{equation}
(- g)\bigl(T^{ik} + _{LL} t^{ik}\bigr) = h^{ikl}_{~~~,l},
\end{equation}
where 
\begin{equation}
h^{ikl} = (-) h^{ilk} = \lambda^{iklm}_{~~~~~~,m}
\end{equation}
and
\begin{equation}
\lambda^{iklm} := \alpha (- g)\bigl(g^{ik} g^{lm} - g^{il}
g^{km}\bigr).
\end{equation}
$\alpha = {1\over 2\beta} = {c^4\over 16\pi G}$.

$_{LL} t^{ik} = _{LL} t^{ki}$ are components of the so--called {\it
Landau--Lifschitz gravitational energy--momentum pseudotensor} and the sum 
\begin{equation}
(- g)\bigl(T^{ik} + _{LL} t^{ik}\bigr) =: _{LL} K^{ik}
\end{equation}
forms the so--called {\it Landau--Lifschitz energy--momentum
complex} \footnote{The complex (25) has worse transformational properties than the Einstein canonical
energy--momentum complex [2,3].} of matter and gravitation. $h^{ikl} = (-)
h^{ilk}$ form the {\it Landau--Lifschitz superpotentials}. 

Then, from the (22)--(23), one can obtain easily 
\begin{eqnarray}
x^i{_{LL}
K^{kl}} - x^k{_{LL} K^{il}}& = & x^i h^{klm}_{~~~~~,m} - x^k
h^{ilm}_{~~~~~,m}\nonumber \\
& = & \bigl(x^i h^{klm} - x^k h^{ilm}\bigr)_{,m} - h^{kli} + h^{ilk},
\end{eqnarray}
or 
\begin{equation}
x^i{_{LL} K^{kl}} - x^k{_{LL} K^{il}} + _{LL} S^{ikl} =
L^{[ik][lm]}_{~~~~~~~~,m},
\end{equation}
where 
\begin{equation}
_{LL} S^{ikl} := h^{ikl} - h^{kil} \not= 0
\end{equation}
and 
\begin{equation}
L^{[ik][lm]} := x^i h^{klm} - x^k h^{ilm}.
\end{equation}

The expression (27) {\it exactly corresponds} to the Bergmann--Thomson
expression (10). Namely, $x^i{_{LL} K^{kl}} - x^k{_{LL} K^{il}}$ gives {\it
orbital part} of the angular momentum density of matter and
gravitation,\footnote{Material part $(-g)\bigl(x^i T^{kl} - x^k
T^{il}\bigr)$ includes material spin also.}  $_{LL} S^{ikl} = h^{ikl} -
h^{kil}$ gives  the non--vanishing {\it spinorial  part} of the gravitational
angular momentum density and $L^{[ik][lm]} = x^i h^{klm} - x^k h^{ilm}$
are {\it superpotentials} for the total angular momentum density of matter and
gravitation.

The total angular momentum density 
\begin{equation}
_{LL} M^{ikl} := x^i _{LL} K^{kl} - x^k _{LL} K^{il} + _{LL} S^{ikl}
\end{equation}
satisfies local conservation laws 
\begin{equation}
_{LL} M^{ikl}_{~~~,l} = 0.
\end{equation}
From (27) one can obtain the following integral expression on the global
angular momentum of an isolated system endowed with an asymptotically
Lorentzian coordinates $(x^0 = ct, ~x^1 = x, ~x^2 = y, ~x^3 = z)$ 
\begin{equation}
_{LL} M^{ik} = {1\over c}\oint\limits_{over ~sphere ~S^2 ~having ~radius
~going~ to ~\infty} L^{[ik][0\alpha]}n_{\alpha} r^2 d\Omega.
\end{equation}

This expression gives the same values of the components $M^{ik} = (-) M^{ki}$
as the Bergmann--Thomson expression (18) gives. Especially, for the Kerr's
spacetime it gives 
\begin{equation}
_{LL} M^{12} = (-) _{LL} M^{21} = mac - {1\over 3}mac = {2\over 3} mac ~;
\end{equation}
other components vanish.

However, in the case we have 
\begin{equation}
_{LL} S^{ikl} = \lambda^{klmi}_{~~~~~,m}
\end{equation}
from which it follows 
\begin{equation}
{_{LL} S^{ikl}}_{,l} = 0.
\end{equation}

The last equality guarantees symmetry of the Landau--Lifschitz energy--momentum
complex: 
\begin{equation}
_{LL} K^{ik} - _{LL} K^{ki} = S^{ikl}_{~~~,l} = 0.
\end{equation}

Landau--Lifschitz, in their book [9], modify the expression (32) on global
angular momentum of an isolated system equipped with an asymptotically
Lorentzian coordinates by using (34). Namely, they {\it subtract}  the
gravitational spin angular density  $_{LL} S^{ikl} = (-) _{LL} S^{kil}$ from
the both sides of the (27) and obtain the {\it new expression} on angular
momentum density of matter and gravitation 
\begin{equation}
x^i {_{LL} K^{kl}} - x^k{_{LL} K^{il}} = \bigl(L^{[ik][lm]} -
\lambda^{klmi}\bigr)_{,m}.
\end{equation}

This new expression includes the {\it total matter angular momentum density},
spinorial  and orbital, but its gravitational part consists of the {\it orbital
gravitational angular momentum density} only.

The new expression (37) leads to the following integrals on global quantities
of an isolated system equipped with an asymptotically Lorentzian coordinates
$(x^0 = ct,~ x^1 = x, ~x^2 = y, ~x^3 = z)$ 
\begin{equation}
M^{ik} = {1\over c}\oint\limits_{over ~sphere ~S^2 ~having~ radius~
R\longrightarrow \infty} \bigl(L^{[ik][0\alpha]} + \lambda^{i0\alpha
k}\bigr)n_{\alpha} r^2d\Omega.
\end{equation}
 
In the case of the Schwarzschild spacetime equipped with an asymptotically
Lorentzian coordinates $(x^0 = ct,x,y,z)$ the expression (38) gives $M^{ik} =
(-) M^{ki} = 0$ and in the case of the Kerr's spacetime it gives 
\begin{equation}
M^{12} = (-) M^{21} = mac - {1\over 3} mac + {1\over 3}mac = mac;
\end{equation}
other components of the $M^{ik}$ vanish.

An analogical modification one can do with the Bergmann--Thomson expression
(18). Namely, by using (11), one can rewrite (10) in the following form

\begin{eqnarray}
x^i{_{BT} K^{jk}} - x^j{_{BT} K^{ik}} & - & \biggl({\alpha\over\sqrt{\vert
g\vert}}\biggr)_{,l}{} g^{[ij][kl]} + \biggl({\alpha
g^{[ij][kl]}\over\sqrt{\vert g\vert}}\biggr)_{,l}\nonumber \\ 
 & = & M^{[ij][kl]}_{~~~~~~~~,l}. 
\end{eqnarray}

By moving the term $\bigl({\alpha g^{[ij][kl]}\over\sqrt{\vert
g\vert}}\bigr)_{,l}$ onto the right hand side of the (40) one gets  a new form
of the Bergmann--Thomson expression (10) with the {\it new angular momentum
complex} $_{BT}{\overline M}^{ikl} = (-) _{BT}{\overline M}^{kil}$ and with the
{\it new superpotentials} ${\overline M}^{[ij][kl]}$.

Namely, we get 
\begin{eqnarray}
x^i{_{BT} K^{jk}} - x^j{_{BT} K^{ik}} & - & \biggl({\alpha\over\sqrt{\vert
g\vert}}\biggr)_{,l}g^{[ij][kl]} \nonumber \\
& = & \biggl(M^{[ij][kl]} - {\alpha\over\sqrt{\vert g\vert}}
g^{[ij][kl]}\biggr)_{,l} =: {\overline M}^{[ij][kl]}_{~~~~~~~~,l}.
\end{eqnarray}

The last expression leads us to the following integrals on global quantities
${\overline M}^{ik} = (-) {\overline M}^{ki}$ in the case of an isolated system
equipped with an asymptotically Lorentzian coordinates $(x^0 = ct,x,y,z)$
\begin{eqnarray}
{\overline M}^{ik} & = & {1\over c}\oint\limits_{over~sphere ~S^2
~having ~radius~R\longrightarrow \infty} \biggl(M^{[ij][0\alpha]} -
\alpha{g^{[ij][0\alpha]}\over\sqrt{\vert g\vert}}\biggr)n_{\alpha} r^2d\Omega
\nonumber \\
 & := & {1\over c}\oint\limits_{over~sphere ~S^2 ~having ~radius ~
R\longrightarrow \infty} {\overline M}^{[ij][0\alpha]} n_{\alpha}r^2
d\Omega.
\end{eqnarray}

The integrals (42) correspond to the Landau--Lifschitz integrals (38) and,
especially, for the Kerr's spacetime they give the same values as the
Landau--Lifschitz integrals (38) give, i.e., they give 
\begin{equation}
{\overline M}^{12} = (-){\overline M}^{21} = mac - {1\over 3}mac + {1\over
3}mac = mac; 
\end{equation}
other components vanish.

So, the Bergmann--Thomson and Landau--Lifschitz integral expressions on global
angular momentum are {\it fully equivalent} in the case of an isolated system
equipped with an asymptotically Lorentzian coordinates.
But we favorize the Bergmann--Thomson expression on angular momentum density
because it has {\it better local transformational properties} in comparison
with the Landau--Lifschitz expression and it is {\it more closely related} to
the canonical energy--momentum complex $_E K_i^{~k} = \sqrt{\vert
g\vert}\bigl(T_i^{~k} + _E t_i^{~k}\bigr)$.

In the case of the Kerr's spacetime one should interpret the value $M^{12} =
(-) M^{21} = {\overline M}^{12} = (-) {\overline M}^{21} = mac$ of the
integrals (38) or (42) as {\it referring  only to the material part of
the angular momentum complex}. It is justified
by the following fact [10]: the integral including the energy--momentum tensor
of matter $T^{ik} = T^{ki}$ only  
\begin{eqnarray}
M^1_{~2} & = & (-) M^{12}  = {1\over c}\int\limits_{x^0 = ct = const} 
{T_i^{~0}\xi^i\sqrt{\vert g\vert}}dxdydz \nonumber \\
 & = & {c^3\over 16\pi G}\oint\limits_{over ~sphere ~S^2 ~having
~radius~R\longrightarrow \infty} 
\biggl(\nabla^0 \xi^{\alpha} - \nabla^{\alpha}\xi^0\biggr)\sqrt{\vert g\vert}
n_{\alpha} r^2d\Omega,
\end{eqnarray}
where $\xi^i = x\delta^i_2 - y\delta^i_1$ are the components of the spatial
Killing vector field ${\vec\xi} = \xi^i\partial_i$ which exists in the case,
gives  
\begin{equation}
M^1_{~2} = (-) M^{12} = (-) mac.
\end{equation}

So, the {\it matter integral} (44) has the same value in the case as the
integrals (38) and (42). In consequence, we conclude that only the
material parts $(-)g\bigl(x^i T^{kl} - x^k T^{il}\bigr),~~\sqrt{\vert
g\vert} \bigl(x^i T^{kl} - x^k T^{il}\bigr)$ of the suitable angular
momentum complex (Landau-Lifschitz or Bergmann-Thomson) give
contribution to the integrals (38) and (42). Gravitational parts {\it give
no contribution} in the used coordinates.

\end{document}